\documentclass{siamltex}

\newtheorem{conjecture}[theorem]{Conjecture}

\newcommand{\prob}{\mbox{Prob}}
\newcommand{\ADV}{{\mathcal B}}
\newcommand{\B}{{\mathcal A}}

\def\binom#1#2{{#1\choose#2}}

\begin{document}

\title{Optimal Bidding Algorithms \\ Against Cheating in
  Multiple-Object Auctions\thanks{A preliminary version appeared in
  {\em Proceedings of the 3rd Annual International Computing and
  Combinatorics Conference}, pages 192--201. 1997.}}

\author{Ming-Yang Kao\thanks{Department of Computer Science, Yale
  University, New Haven, CT 06520, kao-ming-yang@cs.yale.edu.
  Research supported in part by NSF Grant CCR-9531028.}  
\and 
Junfeng Qi\thanks{Department of Economics, Duke University, Durham,
  North Carolina 27708, qijf@econ.duke.edu.}  
\and Lei Tan\thanks{Department of Computer Science, Duke University,
  Durham, North Carolina 27708, lei@cs.duke.edu.}}

\maketitle

\pagestyle{myheadings} \markboth{\sc kao, qi, and tan} {\sc optimal
bidding algorithms for multiple-object auctions}

\begin{abstract}
  This paper studies some basic problems in a multiple-object auction model
  using methodologies from theoretical computer science.  We are especially
  concerned with situations where an adversary bidder knows the bidding
  algorithms of all the other bidders.  In the two-bidder case, we derive
  an optimal randomized bidding algorithm, by which the disadvantaged
  bidder can procure at least half of the auction objects despite the
  adversary's a priori knowledge of his algorithm.  In the general
  $k$-bidder case, if the number of objects is a multiple of $k$, an
  optimal randomized bidding algorithm is found. If the $k-1$ disadvantaged
  bidders employ that same algorithm, each of them can obtain at least
  $1/k$ of the objects regardless of the bidding algorithm the adversary
  uses.  These two algorithms are based on closed-form solutions to certain
  multivariate probability distributions.  In situations where a
  closed-form solution cannot be obtained, we study a restricted class of
  bidding algorithms as an approximation to desired optimal algorithms.
\end{abstract}

\begin{keywords} 
auction theory, bidding algorithms, electronic commerce, automated
negotiation mechanisms, software agents, market-based control
\end{keywords}

\begin{AMS} 
  05A99, 60C05, 68R05, 90A09, 90A12, 90D10, 90D13
\end{AMS}

\section{Introduction}
This paper investigates some basic problems in auction theory.
Broadly speaking, an auction is a market mechanism with explicit or
implicit rules for allocating resources and determining prices on the
basis of bids from market participants \cite{clearwater, HP93, McA87,
Wil92}.  Auctions are frequently used to price various types of
assets.  For instance, the U.S. Treasury raises funds by auctioning
T-bonds and T-notes, while the Department of the Interior sells
mineral rights on federally owned properties via auction.  Economists
are interested in auctions as an efficient way to price and allocate
goods which have no standard market value.  Auctions are believed to
be the simplest and most familiar means of price determination for
multilateral trading without intermediary market makers \cite{HP93,
McA87, Wil92}.

In typical auctions, there are one seller and a group of competing buyers
who bid to possess the auction objects. Procurements describe situations in
which a single buyer wishes to purchase objects from a set of potential
suppliers. There are four basic forms of auctions in use
\cite{HP93,McA87,MW82}.  In an \emph{English auction} or \emph{ascending
  bid auction}, the price of an object is successively raised until only
one bidder remains and wins the object.  In a \emph{Dutch auction}, which
is the converse of an English auction, an initial high price is
subsequently lowered until a bidder accepts the current price.  In a
\emph{first-price sealed-bid auction}, potential buyers submit sealed bids
for an object. The highest bidder is awarded the object and pays the amount
of his bid.  In a \emph{second-price sealed-bid auction}, the highest
bidder wins the object but pays a price equal to the second-highest bid.
While there are many other forms of auctions, these four are of the
greatest interest.

Previous literature on auction theory mainly studied bidding behavior
under the assumption that the objective of bidders is to maximize
expected profits in absence of any budget constraints.  Such work
concentrates on the allocation of a single object to one of many
bidders. Each bidder has a \emph{valuation}, which is his estimate of
the value of the object. In the \emph{independent private valuation}
(IPV) model, each bidder knows his valuation for the object ex
ante. Each bidder's valuation is assumed to be drawn independently
from the same probability distribution. In the \emph{common value}
(CV) model, it is assumed that bidders obtain imperfect estimates of
the value of the object.  The bidders all assign the same value to the
object ex post.  Both models are well studied in auction theory
\cite{Mil81, MW82, Mye81, PS88}.

Very little work in computer science has been conducted on problems
related to auctions. Neither auction mechanisms nor bidding algorithms
have been formally studied.  Nevertheless, computer scientists have
realized the importance of auctions as an efficient method of resource
allocation \cite{clearwater}.  Gagliano~et al.~applied auction
techniques to the allocation of decentralized network
resources~\cite{GFM95}.  Yang~et al.~proposed an auction-based scheme
in which task and resource allocations are determined through
negotiations among system entities~\cite{YBU93}.

Our work investigates some basic issues in the context of automated
negotiation mechanisms which are emerging in electronic commerce and
other applications of software agents for resource allocation.  To
maximize transaction volume and speed, we focus on the auction of
several objects in parallel and propose a {\it multiple-object
auction} model.  This model further differs from the IPV and CV models
in several significant ways.  In this model, each bidder faces a
binding budget constraint which is identical to all the bidders.  Such
constraints can be used to enforce fairness of some form when their
compliance is verifiable.  In electronic transaction environments,
security has been a major concern.  Our model explicitly considers
situations where electronically transmitted information about bids may
be legitimately or illegitimately revealed against the wishes of their
bidders.  In contrast, the IPV and CV models assume that no bidder has
an informational advantage on bids or bidding algorithms over other
bidders~\cite{FT91}.

The assumptions of our model are specified as follows.

\begin{itemize}
\item There are a total of $k$ bidders, $\ADV_1$, $\ADV_2$, \ldots,
  $\ADV_k$, each of whom has the same total resource to devote toward
  winning objects. We normalize this amount to be 1. Assume that $k
  \geq 2$.
\item A total of $n$ objects are auctioned. Assume that $n \geq k$.  Each
  bidder's goal is to maximize the number of objects he wins.  The
  objects are therefore of equal value to a bidder.
\item Each bidder submits a sequence of $n$ bids simultaneously for
  the $n$ objects. Each object is won by the highest bidder at the
  price of his bid.  If $m$ bidders submit the same highest bid for an
  object, each wins the object with probability $1/m$. (Remark: The
  results of our bidding algorithms in \S\ref{c:bida} and
  \S\ref{c:OptimalMultiple} are not affected by the specific
  tie-breaking rules that are used.)  For technical reasons, no zero
  bid is allowed. (Remark: This restriction is only used in
  \S\ref{c:bidb}.)

\begin{figure}
\begin{center}
\vspace{1.5cm}
\begin{picture}(12,12)(0,0)
\setlength{\unitlength}{.2cm}
\put( 0, 5){\circle*{2}}\put(-1,   7){$\ADV_2$}
\put( 0,-5){\circle*{2}}\put(-1,  -8){$\ADV_4$}
\put(-5, 0){\circle*{2}}\put(-9,-0.5){$\ADV_1$}
\put( 5, 0){\circle*{2}}\put( 7,-0.5){$\ADV_3$}
\put( 0,-5){\vector(-1, 1){4.3}}
\put( 0,-5){\vector( 0, 1){9.0}}
\put( 5, 0){\vector(-1,-1){4.3}}
\put( 0, 5){\vector( 1,-1){4.3}}
\end{picture}
\end{center}
\vspace{2cm}
\caption{A graph of information structure} 
\label{fig:info}
\end{figure}
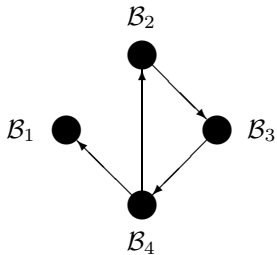

\item Some bidders may know the bidding algorithms of others.  The
  information structure can be characterized by a directed graph in
  which an arc from a bidder $\ADV_i$ to another bidder $\ADV_j$ means
  that $\ADV_i$ knows $\ADV_j$'s algorithm. For instance, in
  Figure~\ref{fig:info}, $\ADV_4$ knows the algorithms of $\ADV_1$ and
  $\ADV_2$; $\ADV_3$ knows $\ADV_4$'s; $\ADV_2$ knows $\ADV_3$'s;
  $\ADV_1$ knows only his own. The bidders all compete
  non-cooperatively.  We assume that each bidder knows the number of
  bidders and that of objects.
\end{itemize}

We analyze the performance of a number of bidding algorithms with which
bidders can assign their bids.  Almost all the bidding algorithms in this
paper are randomized ones.  We first study the case of two bidders, i.e.,
$k=2$, and then extend the results to the case of multiple bidders.  In the
two-bidder case, let $\ADV$ and $\B$ denote the bidders.  We assume that $\B$
knows $\ADV$'s bidding algorithm, while $\ADV$ does not know $\B$'s, i.e., $\ADV$
is a disadvantaged bidder and $\B$ an adversary.  Here, $\B$ is an
oblivious adversary, because although $\B$ knows $\ADV$'s bidding algorithm,
he does not know the outcome of the random choices that $\ADV$ makes.  We
give an optimal randomized bidding algorithm for $\ADV$ by which he can
procure at least one half of the objects despite $\B$'s a priori knowledge
of his bidding algorithm.  The main difficulty with obtaining this optimal
bidding algorithm is finding a closed-form solution to a desired
multivariate probability distribution \cite{BS91, Dal91, Fre51, KS91,
  Sch91}.

We next study the case where there are more than two bidders, and an
adversary bidder knows the bidding algorithms of all the others. If the
number of objects is a multiple of the number of bidders, an optimal
randomized bidding algorithm is found.  If all the disadvantaged bidders
employ that same bidding algorithm, each of them can obtain at least $1/k$
of the objects regardless of the bidding algorithm the adversary uses.
This bidding algorithm is also based on a closed-form solution to a desired
multivariate probability distribution.

When the number of objects is not a multiple of the number of bidders, a
closed-form solution of a desired probability distribution cannot be
obtained. Motivated by this, we study a class of bidding algorithms to
approximate desired optimal algorithms.  A bidding algorithm in this class
computes an initial sequence of bids, and the actual bid sequence is a
random permutation of the initial sequence.

Section~\ref{c:bida} describes the optimal bidding algorithm for the
disadvantaged bidder in the two-bidder case.  In \S\ref{c:OptimalMultiple},
the optimal randomized bidding algorithm from \S\ref{c:bida} is generalized
for the multiple-bidder case. In \S\ref{c:bidb}, a class of bidding
algorithms are introduced to approximate desired optimal algorithms when a
closed-form solution cannot be determined.  Section~\ref{c:con} concludes
the paper.

For brevity, let $W(\ADV_i)$ denote the expected number of objects that
$\ADV_i$ wins with a bidding algorithm that is explicitly or implicitly
specified.

\section{The Two-Bidder Case}
\label{c:bida}
This section studies the two-bidder case. We assume that $\B$ knows $\ADV$'s
bidding algorithm, while $\ADV$ does not know $\B$'s.  We give an optimal
randomized bidding algorithm for $\ADV$ such that $W(\ADV)=n/2$ despite $\B$'s
informational advantage.  Since this problem is a zero-sum game, this bound
of $n/2$ would be straightforward if von Neumann's min-max theorem were
applicable. However, our problem has an infinite pure strategy space, and
it is not immediately clear that the min-max theorem is
applicable~\cite{Bla56,BG54, FS96, FT91,Wil92}.

\subsection{$\ADV$'s Optimal Bidding Algorithm}
\label{s:opt2b}

The following lemma gives an upper bound for the expected number of objects
$\ADV$ can win.
\begin{lemma}\label{l:WAUpperBound}
$W(\ADV) \leq \frac{n}{2}$.
\end{lemma}
\begin{proof}
  Since $\B$ knows $\ADV$'s bidding algorithm, $\B$ can perform at least as
  well as $\ADV$ by employing the same algorithm.  Then this lemma follows
  from the fact that our auction is a zero-sum game.
\end{proof}

Lemma~\ref{l:optfx} describes the marginals of a desired multivariate 
probability distribution with which $\ADV$ can form an optimal
bidding algorithm.

\begin{lemma}
\label{l:optfx}
Assume that $\ADV$ draws his bid sequence $b_1,b_2,\dots,b_n$ from an
$n$-dimensional probability distribution such that each $b_i$ has the same
marginal probability distribution $F_2(b_i)$, where
\begin{equation}
\label{eqn:optfx}
F_2(b_i)= \left\{ \begin{array}{ll}
    \frac{n}{2}{\cdot}b_i    &  b_i \in [0,\frac{2}{n}], \\
    1               &  b_i \in (\frac{2}{n},1],
    \end{array}
\right .
\end{equation}
subject to $\sum b_i =1$. Then, $\B$'s optimal bidding algorithm wins
exactly $n/2$ objects on average.
\end{lemma}
\begin{proof}
  Let $a_1, a_2, \ldots, a_n$ be $\B$'s optimal bids for the $n$ objects,
  respectively. $\B$'s probability of winning the $i$th object is
  $F_2(a_i)$. Since $\ADV$'s bids are within $[0,2/n]$, it is not to $\B$'s
  advantage to bid over $2/n$. Hence $a_i \leq 2/n$ and
  $F_2(a_i)=\frac{n}{2}a_i$.  $\B$'s optimal bids maximize $W(\B)$ as
  follows:
\begin{eqnarray*}
\label{eqn:maxWB}
 \max_{\scriptsize
\begin{array}{c}
\sum a_i=1,\\  
0\leq a_i \leq \frac{2}{n}
\end{array}
}
W(\B)&=&\max_{\scriptsize
\begin{array}{c}
\sum a_i=1,\\  
0\leq a_i \leq \frac{2}{n}
\end{array}
}F_2(a_1)+F_2(a_2)+\cdots+F_2(a_n)\\
&=&\max_{
\scriptsize
\begin{array}{c}
\sum a_i=1 \\
0\leq a_i \leq \frac{2}{n}
\end{array}
} \frac{n}{2}{\cdot}(a_1+\cdots+a_n)= \frac{n}{2}.
\end{eqnarray*}
\end{proof}

Lemma~\ref{l:optsequence} systematically constructs a
bid sequence for $\ADV$ which satisfies the conditions given in
Lemma~\ref{l:optfx}. 
We define two additional functions for Lemma~\ref{l:optsequence}.
Let
\begin{equation}
\label{eqn:sv}
s(v)=\frac{81}{2}{\cdot}\frac{v}{2-3v}.
\end{equation}
Let $h(x,y,z)$ be the function defined on 
\(
 \{(x,y,z)|0\leq x,y,z \leq \frac{1}{3}\}
\)
such that
\begin{equation}
\label{eqn:hxyz}
h(x,y,z)=s(|x-y|+|y-z|+|z-x|) .
\end{equation}

\begin{lemma}
\label{l:hIsDensity}
The function $h(x,y,z)$ is a joint probability density function of $x, y$
and $z$. 
\end{lemma}
\begin{proof}
Note that $h(x,y,z) \geq 0$. To show that $h(x,y,z)$ is a joint probability
density 
function, we need only verify that the integral of $h(x,y,z)$ over
\(  \{(x,y,z)|0\leq x,y,z \leq \frac{1}{3}\} \) is $1$.
Let 
\begin{equation}
\label{eqn:rxy}
r(x,y)=\int_{0}^{\frac{1}{3}}h(x,y,z)dz.
\end{equation}
Consider   the case $x\geq y$. Then
\[
\mbox{if}\  x\geq y \geq z,\ \ \ \ h(x,y,z)=s(2(x-z));
\]
\[
\mbox{if}\  x\geq z \geq y,\ \ \ \ h(x,y,z)=s(2(x-y));
\]
\[
\mbox{if}\  z\geq x \geq y,\ \ \ \ h(x,y,z)=s(2(z-y)).
\]
Hence if $x\geq y$,
\begin{equation}
\label{eqn:lambdaxy}
r(x,y)=\int_0^y s(2(x-z))\,dz+\int_y^x
s(2(x-y))\,dz+\int_x^{1/3}s(2(z-y))\,dz,
\end{equation}
which equals 
\begin{equation}
\label{eqn:rxyreduced}
\frac{9}{2}\left(2\ln(1-3(x-y)) -\ln(3y(1-3x))-\frac{1-6(x-y)}{1-3(x-y)}
\right).
\end{equation}
By symmetry, if $y\geq x$,
\[
r(x,y)=\frac{9}{2}\left(2\ln(1-3(y-x)) -\ln(3x(1-3y))-\frac{1-6(y-x)}{1-3(y-x)}
\right).
\]
It can be verified that 
\[
\int_0^{1/3}\int_0^{1/3}r(x,y)\,dxdy=1.
\]
Thus,
\[
\int_0^{1/3}\!\!\int_0^{1/3}\!\!\int_0^{1/3}
h(x,y,z)\,dxdydz=1.
\]
\end{proof}

\begin{lemma}
\label{l:optsequence}
$\ADV$ can use the following procedure to draw his bids
$b_1,b_2,\ldots,b_n$ such that $\sum b_i =1$ and the marginal probability
distribution of each $b_i$ is as described by~$($\ref{eqn:optfx}$)$.

Case 1: $n=2m$ is even.  $\ADV$ draws $b_1$
from the probability distribution $F_2$ and sets $b_i = b_1$ and
$b_{m+i}=\frac{2}{n}-b_1$ for $i=1,\ldots,m$.  

Case 2: $n=2m+1$ is odd. $\ADV$ draws $b_1$ from $F_2$ and then sets 
$b_i = b_1$  and $b_{m-1+i}=\frac{2}{n}-b_1$ for
$i=1,\ldots,m-1$. For the remaining three bids $b_{2m-1},b_{2m},b_{2m+1},$ 
$\ADV$ draws $(x,y,z)$ according to h in $($\ref{eqn:hxyz}$)$
and sets 
\begin{equation}
\label{eqn:ThreeAs}
 b_{2m-1}=\frac{3}{n}(x-y+\frac{1}{3}), \ \ \ 
 b_{2m}=\frac{3}{n}(y-z+\frac{1}{3}), \ \ \ 
 b_{2m+1}=\frac{3}{n}(z-x+\frac{1}{3}).
\end{equation}
\end{lemma}
\begin{proof}
Note that $\sum_{i=1}^{n}b_i= 1$, whether $n$ is even or odd.

Case 1. This lemma is correct since
if a random variable $X$ is drawn from the uniform probability distribution
 on $[0,\frac{2}{n}]$, then $\frac{2}{n}-X$ has the same
probability distribution. 

Case 2. The proof of Case~1 shows that the marginal probability distribution
of each $b_i$ is $F_2$ for $i=1,\ldots,2m-2$. It  
remains to show that $b_{2m-1},b_{2m},b_{2m+1}$ are also distributed
the same way.
Because these three random variables are symmetric to each other in
(\ref{eqn:ThreeAs}), we only discuss $b_{2m-1}$ in detail. 
Let $t=x-y+\frac{1}{3}$. Since $x$ and $y$ are defined on
$[0,\frac{1}{3}]$, 
$t$ is defined on $[0,\frac{2}{3}]$. We have two cases: $t\in
[0,\frac{1}{3}]$ and  $t\in [\frac{1}{3}, \frac{2}{3}]$.
The two cases are symmetric, and we discuss only the latter. Let $G(t)$
denote  the probability distribution of $t$. Then,
\[
G(t) = 1-{\int\int}_{u-v+\frac{1}{3}\geq t, 0\leq u,v\leq \frac{1}{3}}
r(u,v)\,dvdu=1-\int_{t-1/3}^{1/3}\int_0^{u-t+1/3} 
r(u,v)\,dvdu.
\]
Since $u \geq u-t+1/3$, $r(u,v)$ can take the form of (\ref{eqn:rxyreduced}), 
and we can obtain $G(t)=\frac{3}{2}t$. Since $b_{2m-1} = \frac{3}{n}t$,
$F_2$ is the probability distribution of $b_{2m-1}$.
\end{proof}

\begin{theorem}
\label{t:opta}
The bidding algorithm given in Lemma~\ref{l:optsequence} is optimal for
$\ADV$ and ensures $\ADV$ at least $n/2$  objects in expected terms.
\end{theorem}
\begin{proof}
Lemma~\ref{l:WAUpperBound} gives an upper bound for $W(\ADV)$.
Lemmas~\ref{l:optfx} and~\ref{l:optsequence} give an upper bound for $W(\B)$,
which in turn gives a matching lower bound for $W(\ADV)$ because
$W(\ADV)+W(\B)= n$.
\end{proof}

\subsection{Deriving the Joint Probability Density Function $h(x,y,z)$}
The most difficult step of obtaining the function $h$ is guessing that
$x,y$ and $z$ appear together as $|x-y|+|y-z|+|z-x|$.
It is worthwhile to show the derivation of the function $s$ in 
(\ref{eqn:sv}) that gives the  joint probability density 
function $h(x,y,z)$. 
As in (\ref{eqn:rxy}), let $r(x,y)$ be the probability distribution of
$(x,y)$.
Also let $t=x-y+\frac{1}{3}.$ Since $t=\frac{n}{3} b_{2m-1}$ needs to
be uniformly distributed over $[0,\frac{2}{3}]$, we need to have 
\begin{equation}
\label{eqn:NeedToHave}
 1-\int_{t-1/3}^{1/3}\int_0^{u-t+1/3}
r(u,v)\,dvdu = \frac{3}{2}t,
\ \ \mbox{for all\ } t\in [\frac{1}{3},\frac{2}{3}],
\end{equation}
and
\[
 \int_{0}^{t}\int_{u-t+1/3}^{1/3}r(u,v)\,dvdu = \frac{3}{2}t,
\ \ \mbox{for all\ } t\in [0,\frac{1}{3}].
\]
These two cases are symmetric, and we only discuss the case
given by (\ref{eqn:NeedToHave}) in detail. 
For notational simplicity, let 
\[
s(2v)=q(v), \ \ \int_{}^{u} q(v)\,dv=p(u),\ \
r_2(x,y)=\frac{\partial{r(x,y)}}{\partial{y}}.
\]
Differentiating (\ref{eqn:NeedToHave}) with respect to $t$ twice, we
 obtain 
\begin{equation}
\label{eqn:twodir}
\int_{t-1/3}^{1/3}r_2(u,u-t+\frac{1}{3})\,du +
r(t-\frac{1}{3},0)=0.
\end{equation}
Since $x \geq y$ in (\ref{eqn:NeedToHave}), the following is derived
from (\ref{eqn:lambdaxy}):   
\[
r_2(u,v)=-(u-v)q'(u-v)-q(\frac{1}{3}-v)+q(u-v).
\]
Then,
\begin{eqnarray}
\label{eqn:sub1r2}
& & \int_{t-1/3}^{1/3}r_2(u,u-t+\frac{1}{3})\,du 
\\  & = & -(t-\frac{1}{3})
(\frac{2}{3}-t)q'(t-\frac{1}{3})+p(t-\frac{1}{3})-p(\frac{1}{3}) 
+q(t-\frac{1}{3})(\frac{2}{3}-t). 
\nonumber
\end{eqnarray}
We obtain from (\ref{eqn:lambdaxy}),
\begin{equation}
\label{eqn:sub2rt}
r(t-\frac{1}{3},0)=(t-\frac{1}{3})q(t-\frac{1}{3})+p(\frac{1}{3})-p(t-
\frac{1}{3}).
\end{equation}
Setting $w=t-\frac{1}{3}$, we can derive the following 
differential equation from (\ref{eqn:twodir}), (\ref{eqn:sub1r2}) and
(\ref{eqn:sub2rt}):
\[w(1-3w)q'(w)=q(w).\]
The solution to the differential equation is 
\[
q(w)=c\frac{3w}{1-3w},
\]
where $c$ is a constant. Therefore 
\[
s(v)=c\frac{3v}{2-3v}.
\]
Since $h(x,y,z)$ is a probability density function for $(x, y, z)$,
$c$ is set to $\frac{27}{2}$ to satisfy
\[
\int_0^{1/3}\!\!\int_0^{1/3}\!\!\int_0^{1/3}
h(x,y,z)\,dxdydz=1.
\]

\section{The Multiple-Bidder Case}
\label{c:OptimalMultiple}
This section generalizes the results in \S\ref{c:bida} to give 
an optimal randomized bidding algorithm for the case of multiple bidders. 
We assume that the bidding algorithms of $k-1$ bidders are known
to a single adversary bidder $\B$. If all the $k-1$ disadvantaged bidders
employ our bidding algorithm, each of them wins at least a fraction $1/k$ of the
objects regardless of the bidding algorithm the 
adversary uses.

\begin{lemma}
\label{l:kbiddermarginal}
Assume that each of the $k-1$ disadvantaged bidders independently 
draws his bid sequence
$b_1, b_2, \dots, b_n$ from an $n$-dimensional probability distribution
such that each $b_i$ has the same marginal probability distribution
$F_k(b_i)$, where
\begin{equation}
\label{eqn:kFx}
F_k(b_i)= \left\{ 
  \begin{array}{ll}
    {\left(\frac{n}{k}{\cdot}b_i\right)}^{\frac{1}{k-1}} &
    \mbox{if $b_i\in [0,\frac{k}{n}]$,} \\ 
    1              &  \mbox{if $b_i\in (\frac{k}{n},1]$, }
  \end{array}
\right .
\end{equation}
subject to $\sum b_i =1$. Then, $W(\B)$ is at most $n/k$. 
\end{lemma}
\begin{proof}
  Let $b_{i,j}$ denote the bid on the $i$-th object of the $j$th
  disadvantaged bidder.  Let $a_i$ be $\B$'s bid on the $i$-th object.
  Because the bids of the $k-1$ disadvantaged bidders are within $[0,k/n]$,
  $\B$ has no incentive to bid over $k/n$. Thus, $a_i \leq k/n$, and
  $F(a_i) = {\left(\frac{n}{k}{\cdot}a_i\right)}^{\frac{1}{k-1}}$.  Since
  bids from different disadvantaged bidders are independent,
\begin{eqnarray*}
  & & \prob\{a_i \mbox{wins the $i$-th object}\}
\\ &=& \prob\{b_{i,1}\leq a_i\}{\cdot}\prob\{b_{i,2}\leq a_i\}\cdots
\prob\{b_{i,k-1}\leq a_i\}
\\ &=& {\left( F_k(a_i) \right)}^{k-1}
\\ &=& \frac{n}{k}{\cdot}a_i.
\end{eqnarray*}
From the fact that  $\sum a_i \leq 1$,  $\B$ wins exactly
$n/k$ objects on average. 
\end{proof}

It appears quite difficult to find a closed-form solution to a joint
probability distribution whose marginals are as described by
(\ref{eqn:kFx}).

\begin{conjecture}
\label{con:dist}
There exists an $n$-dimensional joint probability distribution such that its
marginal probability distribution of every component is as described by
$($\ref{eqn:kFx}$)$, while the components from all dimensions sum to 1.  
\end{conjecture}

For $k=2$, this conjecture has been proved in \S\ref{c:bida}. If $n$ is a
multiple of $k$, we prove this conjecture as follows. 
Let
\[
e(b_1,b_2,\dots,b_k)= \left\{ \begin{array}{ll}
    (b_1b_2\cdots b_k)^{\frac{1}{k-1}-1}  & \mbox{$b_1+b_2+\cdots+b_k
      =1$, $b_i > 0$;}\\  
    0 & \mbox{otherwise.}
  \end{array}
\right .
\]
Let
\[
\alpha =  \int_{b_1+\dots+b_k=1} e(b_1,b_2,\dots,b_k) db_1 db_2\cdots db_{k-1}.
\]
Normalizing $e$ using $\alpha$, we have 
\begin{equation}
\label{eqn:kbidderjoint}
g(b_1,b_2,\dots,b_k)= \left\{ 
\begin{array}{ll}
\frac{{(b_1b_2\cdots b_k)}^{\frac{1}{k-1}-1}}{\alpha}  &
\mbox{$b_1+b_2+\cdots+b_k   =1$, $b_i>0$; }\\  
0 & \mbox{otherwise.}
    \end{array}
\right .
\end{equation}
With this normalization, $g$ is a probability density function of
$(b_1,b_2,\dots, b_k)$.
For example, if $n=k=3$, the probability density function shown in
(\ref{eqn:kbidderjoint}) is
\[
g(b_1,b_2,b_3)= \left\{ \begin{array}{ll}
\frac{1}{2\pi\sqrt{b_1 b_2 b_3}} & \mbox{ 
  $b_1+b_2+b_3 =1, b_i>0$;} \\
0 & \mbox{otherwise}.
    \end{array}
\right .
\]

The following lemma proves Conjecture~\ref{con:dist} for the case $n=k$. 
\begin{lemma}
\label{l:nkequal}
If $n=k$ and the bid sequence $b_1, b_2, \ldots, b_n$ is drawn from the
$n$-dimensional joint probability distribution in
$($\ref{eqn:kbidderjoint}$)$, then the marginal probability distribution
for each $b_i$ is as described by $($\ref{eqn:kFx}$)$. 
\end{lemma}
\begin{proof}
Because $b_1,b_2,\ldots,b_k$ are symmetric for $g$,
we need only show that the probability distribution of $b_k$ is as described in
 (\ref{eqn:kFx}). Let 
\[
b_{i} = (1-b_k) u_{k-i}, i =2,\ldots, k-1.
\]
Then
\[
d b_{i} = (1-b_k) du_{k-i}.
\]
Let
\[
\alpha' =  \int_{u_1+\dots+u_{k-1}=1}
{(u_1u_2{\dots}u_{k-1})}^{\frac{1}{k-1}-1} du_{k-2}  du_{k-3}{\cdots}du_1.
\]
Note that $\alpha = (k-1){\cdot}\alpha'$. The probability distribution of
$b_k$ equals 
\begin{eqnarray*}
& & \int_{
\scriptsize
\begin{array}{l}
0 \leq w \leq b_k \\
b_1 + b_2 + \cdots + b_{k-1}+w =1
\end{array}
} g(b_1, \cdots, b_{k-1}, w) db_2 \cdots db_{k-1} dw \\
&=& \frac{1}{\alpha}{\cdot}\int_{0}^{b_k}\int_{0}^{1-w}\!\!\int_{0}^{1-w-b_{k-1}}\!\!
\cdots \int_{0}^{1-w-{\cdots}-b_3}  
\\ 
&& \hspace{0.5in} {\left( (1-b_2-\cdots-w)b_2 b_3{\cdots}w
  \right)}^{\frac{1}{k-1}-1} db_2\ db_3{\cdots}db_{k-1} dw\\ 
 &=&\frac{1}{\alpha}{\cdot}\int_{0}^{b_k}w^{\frac{1}{k-1}-1}
 \int_{0}^{1}\!\!\int_{0}^{1-u_{1}}\!\!{\cdots}\int_{0}^{1-u_1-\cdots  u_{k-3}}
 \\
 && \hspace{0.5in}{\left( u_1 u_2\cdots u_{k-2}
     (1-u_1-\cdots-u_{k-2})\right) }^{\frac{1}{k-1}-1}
 du_{k-2}{\cdots}du_2du_1dw\\ 
&=& \frac{1}{\alpha}{\cdot}\int_{0}^{b_k}\alpha'w^{\frac{1}{k-1}-1} dw \\
&=&  \frac{\alpha'}{\alpha}(k-1)a^{\frac{1}{k-1}} \\
&=&F_k(b_k). 
\end{eqnarray*}
\end{proof}

The following lemma extends Lemma~\ref{l:nkequal} to the case $n=k{\cdot}m$
for some integer.

\begin{lemma}
\label{l:optimalkm}
If $n=k{\cdot}m$ for some integer $m$, there exists a procedure to generate
a bid sequence $b_1, b_2, \dots, b_n$ such that the probability
distribution for each $b_i$ can be described by $($\ref{eqn:kFx}$)$, and
the bids $b_i$ sum to 1
\end{lemma}
\begin{proof}
If $m=1$, the lemma is the same as Lemma~\ref{l:nkequal}. If $m>1$, we
 divide the objects into $m$ groups of $k$ objects each and employ
Lemma~\ref{l:nkequal} to obtain bids for the first group. We then
set the bids for the other $m-1$ groups to the corresponding
bids for the first group. We scale every bid by a factor of $\frac{1}{m}$
so that the bids sum to $1$.
This gives the desired probability distribution.
\end{proof}

\begin{theorem}
  If $n=k{\cdot}m$ for some integer $m$, and the disadvantaged bidders all
  employ the bidding algorithm characterized by Lemma~\ref{l:optimalkm},
  then each can obtain at least $n/k$ objects in expected terms, which is
  optimal.
\end{theorem}
\begin{proof}
  From Lemmas~\ref{l:kbiddermarginal} and~\ref{l:optimalkm} and the fact
  that our game is a zero-sum game, the $k-1$ disadvantaged bidders win
  $\frac{k-1}{k}{\cdot}n$ objects in total. Since they all use the same
  bidding algorithm, by symmetry, each of them wins $n/k$ objects. This
  upper bound of $n/k$ is also a lower bound since the adversary can always
  win at least $n/k$ objects by employing the same bidding algorithm as the
  disadvantaged bidders.
\end{proof}

\section{Position-Randomized Bidding Algorithms}
\label{c:bidb}
In \S\ref{c:OptimalMultiple}, an optimal randomized bidding algorithm for
the bidders with informational disadvantage is derived for the case where
the number of objects is a multiple of that of bidders.  This algorithm is
based on a closed-form solution to a desired multivariate probability
distribution. If $n$ is not a multiple of $k$, a closed-form solution
cannot be obtained with our current techniques. Motivated by this, we
consider situations where all the bidders are restricted to a class of
bidding algorithms called \emph{position-randomized bidding algorithms}.  A
position-randomized bidding algorithm consists of two steps.  Step~1
deterministically selects an initial sequence of $n$ bids. Step~2 permutes
the sequence. The $i$-th element of the final sequence is the actual bid
for the $i$-th object.  As in \S\ref{c:OptimalMultiple}, we assume that all
the disadvantaged bidders adopt an identical bid sequence at Step~1 and the
same probability distribution at Step~2.  A position-randomized bidding
algorithm can be considered as an approximation to optimal bidding
algorithms desired for resolving Conjecture~\ref{con:dist} in
\S\ref{c:OptimalMultiple}.

The next lemma examines how probability distributions chosen at Step 2
affect the expected numbers of objects bidders win.
\begin{lemma}\label{l:Nonuniform}
  For a given initial bid sequence $a_1,a_2,\ldots,a_n$ of $\B$ and a given
  initial bid sequence $b_1,b_2,\ldots,b_n$ of the disadvantaged bidders,
\begin{itemize}
\item $W_1$ denotes the expected number of objects $\B$ wins using the
  uniform probability distribution while the disadvantaged bidders may use
  any arbitrary probability distribution;
\item $W_2$ denotes the expected number of objects $\B$ wins without
  permuting his initial bid sequence while the disadvantaged bidders employ
  the uniform probability distribution;
\item $W_3$ denotes the expected number of objects $\B$ wins using any
  given probability distribution while the disadvantaged bidders employ the
  uniform probability distribution.
\end{itemize}
If $a_1,a_2,\ldots,a_n$ are all different from $b_1,b_2,\ldots,b_n$, then
$W_1 \geq W_2 = W_3$.
\end{lemma}
\begin{proof}
  For each $a_i$, 
\begin{itemize}
\item $W_{1,i}$ denotes the expected number of objects $a_i$ wins if $\B$
  uses the uniform probability distribution while the disadvantaged bidders
  may use any arbitrary probability distribution;
\item $W_{2,i}$ denotes the expected number of objects $a_i$ wins if $\B$
  does not permute his initial bid sequence and the disadvantaged bidders
  employ the uniform probability distribution; 
\item $W_{3,i}$ denotes the expected number of objects $a_i$ wins if $\B$
  uses a given probability distribution and the disadvantaged bidders
  employ the uniform probability distribution.
\end{itemize}
Since $W_j = W_{j,1}+\cdots+W_{j,n}$ for $j \in \{1, 2, 3\}$, it suffices to
prove that $W_{1,i} \geq W_{2,i} = W_{3,i}$.  Without loss of generality,
assume that $b_1 \leq b_2 \leq \cdots \leq b_n$.  Let $p$ be the largest
index such that $b_p < a_i$; if no such $b_p$ exists, let $p = 0$.  Since
$a_i < b_j$ for $j = p+1,\ldots,n$,
  \[W_{2,i} = \left(\frac{p}{n}\right)^{k-1}.\]
  To calculate $W_{3,i}$, let $Q_{q,r}$ be the probability that $\B$ places
  $a_q$ on the $r$-th object.  Then,
  \begin{eqnarray*}
    W_{3,i} & = & \sum_{r=1}^{n}\prob\{a_i\ \mbox{wins the $r$-th object}\}
    \\ 
    & = & \sum_{r=1}^{n}Q_{i,r}{\cdot}\left(\frac{p}{n}\right)^{k-1}.
  \end{eqnarray*}
  Since $\sum_{r=1}^n Q_{i,r}=1$, \[W_{2,i}=W_{3,i}.\] To calculate
  $W_{1,i}$, let $P_{q,r}$ be the probability that a disadvantaged bidder
  places $b_q$ on the $r$-th object.  Then,
  \begin{eqnarray*}
    W_{1,i} & = & \sum_{r=1}^{n}\frac{1}{n}{\cdot}\prob\{a_i\ \mbox{wins
      the $r$-th object}\}
  \\ & = &
  \sum_{r=1}^{n}\frac{1}{n}{\cdot}{(P_{1,r}+P_{2,r}+\cdots+P_{p,r})}^{k-1}.
  \end{eqnarray*}
  Since $\sum_{r=1}^n P_{q,r}=1$ for each $q$, by H\"{o}del's inequality,
  \[W_{1,i} \geq W_{2,i}.\]
\end{proof}

Since $W_1 \geq W_3$ in Lemm~\ref{l:Nonuniform}, the disadvantaged
bidders should always use the uniform probability distribution at Step
2.  Since $W_2 = W_3$, we may assume that $\B$ does not permute his
initial bid sequence whenever the disadvantaged bidders use the
uniform probability distribution.  We next use
Lemma~\ref{l:Nonuniform} to derive a lower bound for the expected
number of objects $\B$ can win.  Let
\begin{eqnarray*}
\epsilon & = &  \mbox{a positive infinitesimal amount};
\\
\beta & = & \sum_{i=1}^{n} i^{k-1};
\\
c_i & = & \frac{i^{k-1}}{\beta};
\\
{E} & = & \{ c_0, c_1, c_2, \dots, c_n \};
\\
{D} & = & \{ \epsilon, c_2+\epsilon, c_3+\epsilon, \dots, c_{n}+\epsilon\}.
\end{eqnarray*}

\begin{lemma}\label{l:KBidderWinRatio}
  $\B$ can win at least $\frac{\beta-1}{n^{k-1}}$ objects on average for
  any given initial bid sequence and probability distribution employed by
  the disadvantaged bidders.
\end{lemma}
\begin{proof}
  Given an initial bid sequence $b_1 \leq b_2 \leq \cdots \leq b_n$ of the
  $k-1$ disadvantaged bidders, $\B$ chooses his initial bid sequence to be
  $b_1-(n-1)\epsilon, b_2+\epsilon, \ldots, b_n+\epsilon$.  Since $\B$'s
  bids are different from $b_1, b_2, \ldots, b_n$, in light of
  Lemma~\ref{l:Nonuniform}, we may assume that the disadvantaged bidders
  permute their bids with the uniform probability distribution.
  Consequently, the expected number of objects won by $\B$ is as desired.
\end{proof}

We next prove a matching upper bound for the expected number of objects
$\B$ can win.
\begin{lemma}
\label{l:prk2}
If the disadvantaged bidders employ $c_1,c_2,\ldots,c_n$ as their initial
bid sequence and permute it with the uniform probability distribution, then
$\B$ has an optimal initial bid sequence $a'_1, a'_2, \dots, a'_n$ such
that $a'_i \in {D}$ for all $i$.
\end{lemma}
\begin{proof} 
  Given an optimal initial bid sequence $a_1, a_2, \ldots, a_n$ of $\B$, we
  show that this sequence can be transformed into a desired sequence $a'_1,
  a'_2, \ldots, a'_n$ without decreasing $W(\B)$.  Let $m$ be the number of
  $\B$'s bids that are in ${E}$.  There are three cases.

  Case~1: $m=0$.  For each $a_i$, let $a'_{i} = c_j+\epsilon$ where
  $j$ is the biggest index such that $c_j < a_i$. Then the expected number
  of objects  won by $a'_1, a'_2,\dots,a'_n$ is the same as that of $a_1,
  a_2, \dots, 
  a_n$, and the new sequence is as desired.

  Case~2: $m=1$. This case is impossible since $\B$ can increase $W(\B)$ by
  decreasing one of his bids outside ${E}$ by $\epsilon$ and increasing the
  one that is in ${E}$ by $\epsilon$.

  Case~3: $m\geq 2$.  Without loss of generality, let $a_1, a_2, \dots,
  a_m$ be $\B$'s $m$ bids in ${E}$ in the increasing order. We first
  decrease $a_1$ by $(m-1)\epsilon$ and increase $a_j$ by $\epsilon$ for
  $j=2,\dots, m$.  As shown below, this adjustment never decreases $W(\B)$.
  Then, since $\B$'s adjusted bids are not in ${E}$, his new initial bid
  sequence can be further transformed into a desired sequence as in Case~1.
  Let $w_1$ be the decreased amount of $W(\B)$ resulted from decreasing
  $a_1$.  Let $w_j$ be the increased amount of $W(\B)$ resulted from
  increasing $a_j$ for $j=2,\cdots,m$. We need to show that
  $-w_1+w_2+\cdots+w_m \geq 0$. It suffices to prove that $w_2 - w_1 \geq
  0$.  Let ${\#}_p$ denote the expected number of objects $a_j$ wins if
  $a_j = c_p$.  Then,
\begin{eqnarray*}
\#_p & = & \sum_{i=0}^{k-1} \frac{1}{i+1}{\cdot}\prob\{\mbox{$a_j$ ties
  with $i$ disadvantaged bidders and beats the others}\}
\\ 
& = & \sum_{i=0}^{k-1} \frac{1}{i+1}\binom{k-1}{i} \left(\frac{
  1}{n}\right)^i\left(\frac{p-1}{n}\right)^{k-1-i}
\\ 
& = & \sum_{i=0}^{k-1} \frac{1}{i+1}\binom{k-1}{i}
\frac{(p-1)^{k-1-i}}{n^{k-1}} 
\\ 
& = &
\left(\frac{p-1}{n}\right)^{k-1}{\cdot}\frac{1}{k}{\cdot}\left(
\left(\frac{1}{p-1}\right)^{k}-1\right).    
\end{eqnarray*}
Assume that $a_1 = c_{{q}}$ and $a_2 = c_{{r}}$. Then,
$w_1={\#}_{{q}}-(\frac{{q}-1}{n})^{k-1}$ and
$w_2=(\frac{{{r}}}{n})^{k-1}-{\#}_{{r}}$.  Note that $w_2$ increases with
${r}$.  Since ${q} \leq {r}$, $w_2$ is minimized when $a_1=a_2$ and thus
$q=r$.  Consequently,
\begin{eqnarray*} 
  w_2-w_1 & \geq &
  \left(\frac{{q}}{n}\right)^{k-1}-
\#_{{q}}-\#_{{q}}+\left(\frac{{q}-1}{n}\right)^{k-1}
\\ 
& = & \left(\frac{{q}}{n}\right)^{k-1} + \left(\frac{{q}-1}{n}\right)^{k-1} -
2{\cdot}\sum_{i=0}^{k-1} \frac{1}{i+1}\binom{k-1}{i}
\frac{(q-1)^{k-1-i}}{n^{k-1}} 
\\ 
& = & \left(\frac{{q}}{n}\right)^{k-1} - \left(\frac{{q}-1}{n}\right)^{k-1} -
2{\cdot}\sum_{i=1}^{k-1} \frac{1}{i+1}\binom{k-1}{i}
\frac{(q-1)^{k-1-i}}{n^{k-1}} 
\\ 
\ & = &
\sum_{i=1}^{k-1}
\binom{k-1}{i}\left(\frac{q-1}{n}\right)^i\left(\frac{1}{n}\right)^{k-1-i}-
2{\cdot}\sum_{i=1}^{k-1} \frac{1}{i+1}\binom{k-1}{i}
\frac{(q-1)^{k-1-i}}{n^{k-1}} 
\\ 
& = & \sum_{i=1}^{k-1}
\left(1-\frac{2}{i+1}\right)\binom{k-1}{i}\frac{(q-1)^{k-1-i}}{n^{k-1}}
\\ 
\ & \geq & 0.
\end{eqnarray*}
\end{proof}

\begin{lemma}
\label{l:optseqk}
If the $k-1$ disadvantaged bidders all employ $c_1,c_2,\ldots,c_n$ as their
initial bid sequence and permute it with the uniform probability
distribution, then $\B$ can win at most $\frac{\beta-1}{n^{k-1}}$ objects
on average.
\end{lemma}
\begin{proof} 
  From Lemma~\ref{l:prk2}, $\B$ has an optimal initial bid sequence
  $a'_1,a'_2, \ldots, a'_n$, such that for all $j$, $a'_j \in {D}$. If
  $a'_j = \epsilon$, then it cannot win any object. If $a'_j =
  c_i+\epsilon$, then it can win $(\frac{i}{n})^{k-1}$ objects on average.
  The unit price $\B$ pays for these objects is strictly greater than
\[
\frac{\frac{i^{k-1}}{\beta}}{(\frac{i}{n})^{k-1}}=
\frac{n^{k-1}}{\beta}.
\]
Since the expected number of objects won by such $a'_j$ is an integral
multiple of $\frac{1}{n^{k-1}}$, $W(\B)=m{\cdot}\frac{1}{n^{k-1}}$ for some
integer $m$, and
\[
m{\cdot}\frac{1}{n^{k-1}}{\cdot}\frac{n^{k-1}}{\beta} < 1.
\]
Since $m$ is an integer, $m \leq \beta-1$ and thus $ W(\B) \leq
\frac{\beta-1}{n^{k-1}}$.
\end{proof}

\begin{theorem}
  If the disadvantaged bidders all employ $c_1,c_2,\ldots,c_n$ as their
  initial bid sequence and permute it with the uniform probability
  distribution, then each of them can win at least $1/k$ of
  $n-\frac{\beta-1}{n^{k-1}}$ objects on average, which is optimal.
\end{theorem}
\begin{proof}
  By Lemma~\ref{l:optseqk}, $\B$ wins at most $\frac{\beta-1}{n^{k-1}}$
  objects on average.  By Lemma~\ref{l:KBidderWinRatio}, this upper bound
  is also the lower bound of the expected number of objects $\B$ can win.
  Then this theorem follows from the fact that our auction is a zero-sum
  game.
\end{proof}

\section{Extensions and Open Problems}
\label{c:con}
This paper leaves several problems unsolved.
Section~\ref{c:OptimalMultiple} still lacks an optimal randomized bidding
algorithm for the disadvantaged bidders when $n$ is not a multiple of $k$.
In \S\ref{c:bidb}, if zero bids are allowed, the initial bid sequence
$c_1,\ldots,c_n$ is no longer optimal for the disadvantaged bidders. In
general, if disadvantaged bidders do not use identical bidding algorithms,
it is not even clear what an optimal bidding algorithm should mean,
especially for a more complicated information structure than discussed in
this paper.

Our model can be extended to study sequential bidding. The bidders submit
sealed bids for an object. Once that object is sold, the next object is
auctioned the same way until all the objects are sold.  For the case where
$n$ is a multiple of $k$, an optimal sequential bidding algorithm is
described in the following lemma.

\begin{lemma}
  If $n$ is a multiple of $k$ and the objects are auctioned sequentially,
  then a bidder can obtain $n/k$ objects by bidding $k/n$ on every object
  until his budget is exhausted.
\end{lemma}
\begin{proof}
  Assume that $\ADV_i$ employs this bidding algorithm.  From his budget
  constraint, he wins at most $n/k$ objects. This upper bound is also a
  lower bound.  To prove this claim by contradiction, assume that $\ADV_i$
  wins fewer than $n/k$ objects and thus does not exhaust all his budget.
  Then, the total number of objects won by the other bidders exceeds
  $\frac{k-1}{k}{\cdot}n$. Because $n$ is a multiple of $k$ and $\ADV_i$ has
  not exhaust his budget, every object's winning bid must be at least
  $k/n$.  Therefore, the total of the winning bids of the other bidders
  exceeds $k-1$.  Since this contradicts the budget constraint, $\ADV_i$ can
  win at least $n/k$ objects.
\end{proof}

Our model can also be extended to the case where the objects may have
distinct values.  In a general setting, the objects are divided into $m$
groups. Let $n_i$ denote the number of objects in the $i$-th group, which
may be any positive real number. The bidders are asked to submit bids for
the $m$ groups simultaneously. Whoever bids the highest for a group obtains
all the objects in that group subject to the same tie-breaking rule.  An
$m$-group auction is equivalent to an auction of $m$ objects with distinct
values where $n_i$ is the value of the $i$-th group.  As before, assume
that an adversary bidder $\B$ knows the bidding algorithms of the other
$k-1$ bidders, and all those disadvantaged bidders employ the same bidding
algorithm.

\begin{lemma}\label{l:mgroupmarginal}
  Assume that each disadvantaged bidder bids $n_i{\cdot}b_i$ for the $i$-th
  group where $b_1, b_2, \dots, b_m$ are drawn from an $m$-dimensional
  probability distribution such that the marginal probability distribution
  of each $b_i$ is $F_k$ subject to
  $n_1{\cdot}b_1+{\cdots}+n_m{\cdot}b_m=1$. Then the optimal expected
  number of objects won by $\B$ is $n/k$.
\end{lemma}
\begin{proof}
  Let $n_i{\cdot}b_{i,j}$ denote the bid on the $i$-th object by the $j$-th
  disadvantaged bidder.  Let $n_i{\cdot}a_i$ be $\B$'s bid on the $i$-th
  object.  Because $b_1,b_2,\dots,b_m \in [0,k/n]$, $\B$ has no incentive
  to set $a_i$ greater than $k/n$. Thus, $a_i \leq k/n$ and $F_k(a_i) =
  {\left(\frac{n}{k}{\cdot}a_i\right)}^{\frac{1}{k-1}}$.  Since bids from
  different disadvantaged bidders are independent,
\begin{eqnarray*}
 & & \prob\{n_i{\cdot}a_i\ \mbox{wins the $i$-th object}\} 
\\ &=& \prob\{b_{i,1}\leq a_i\}{\cdot}\prob\{b_{i,2}\leq a_i\}\cdots
\prob\{b_{i,k-1}\leq a_i\}
\\ &=& {\left( F_k(a_i) \right)}^{k-1} 
\\ &=& \frac{n}{k}{\cdot}a_i.
\end{eqnarray*}
$\B$ maximizes $W(\B)$ as follows:
\begin{eqnarray*}
& & 
\max_{
\scriptsize
\begin{array}{c}
  \sum n_i{\cdot}a_i =1\\ 1 \leq a_i \leq \frac{k}{n}
\end{array}
}
W(\B)  
\\ & = & 
\max_{\scriptsize
\begin{array}{c}
  \sum n_i{\cdot}a_i = 1\\ 1 \leq a_i \leq \frac{k}{n}
\end{array}
}n_1{\cdot}(\frac{n}{k}{\cdot}a_1) + n_2{\cdot}(\frac{n}{k}{\cdot}a_2)
+\cdots+ n_m{\cdot}(\frac{n}{k}{\cdot}a_m)
\\
& = & \frac{n}{k}.
\end{eqnarray*}
\end{proof}

\begin{conjecture}
\label{con:group}
There exists an $m$-dimensional probability distribution for
$(b_1,b_2,\dots,b_m)$ subject to the constraint
$n_1{\cdot}b_1+n_2{\cdot}b_2+\cdots+n_m{\cdot}b_m=1$ such that the marginal
probability distribution of each $b_i$ is as described by
$($\ref{eqn:kFx}$)$.
\end{conjecture}

{\it Remark.} This conjecture can be reduced to the case $m=2$ or $3$.

We conclude the paper with two research directions. One is to consider
general information structures as specified by arbitrary directed
graphs.  The other is to investigate more general budget constraints
beyond the homogeneous one of this paper.  It would be of significance
to design bidding algorithms that can optimally or approximately
achieve game-theoretic equilibria in meaningful combinations of these
two directions.

\section*{Acknowledgements}
We are indebted to Phil Long, Kasturi Varadarajan, and Professor
Dennis Yang at the Economics Department of Duke University for very
helpful comments. We also wish to thank anonymous referees for
contagious enthusiasm towards this work and unusually thoughtful
comments and detailed suggestions.


\end{document}